\documentclass[11pt]{report}
\usepackage[proposal]{USCthesis}

\newcommand{\note}[1]{}

\usepackage{times}
\usepackage{helvet}
\usepackage{courier}
\usepackage{bm}

\usepackage[latin9]{inputenc}
\usepackage{array}
\usepackage{verbatim}
\usepackage{float}
\usepackage{amsmath}
\usepackage{amssymb}
\usepackage{graphicx}
\usepackage{epsfig}
\usepackage{subfigure}
\usepackage{epstopdf}
\usepackage{algorithm}
\usepackage{algorithmic}
\usepackage{wrapfig}
\usepackage{color}
\usepackage{wrapfig}
\usepackage{lipsum}
\usepackage{booktabs}
\usepackage{hyperref}
\usepackage{url}

\DeclareMathOperator*{\argmax}{argmax}

\DeclareMathOperator{\E}{\mathbb{E}}

\newcommand{\tuple}[1]{\ensuremath{\left \langle #1 \right \rangle }}

\newtheorem{problem}{\hskip\parindent\bf{Problem}}
\newtheorem{theorem}{\hskip\parindent\bf{Theorem}}

\def\theselectedpubs#1{
  \chapter*{Key Publications}
  \list{[\arabic{enumi}]}{
    \settowidth\labelwidth{[#1]}
    \leftmargin\labelwidth   \advance\leftmargin\labelsep
    \ifopenbib
      \listparindent -1.5em
      \advance\leftmargin-\listparindent
      \itemindent\listparindent
      \parsep 0pt%
    \fi
    \usecounter{enumi}}%
  \ifopenbib
    \def\newblock{\par}
    \let\\=\@centercr
    \@rightskip\@flushglue   \rightskip\@rightskip
    \leftskip\z@
  \else
    \def\newblock{\hskip .11em plus .33em minus -.07em}%
  \fi
  \sloppy
  \sfcode`\.=1000\relax}

\floatstyle{ruled}
\newfloat{algorithm}{tbp}{loa}
\providecommand{\algorithmname}{Algorithm}
\floatname{algorithm}{\protect\algorithmname}



\pagetop{1.1 true in}       
\pageleft{1.6 true in}      
\pageheight{8.5 true in}    
\pagewidth{5.85 true in}    
\pagemargin{1.0}    

\usepackage{checkend}

\begin{document}

\title{Influence Maximization for Social Good: Use of Social Networks in Low Resource Communities}
\author{Amulya Yadav}
\majorfield{Computer Science}

\committee{Milind Tambe & (Chairperson) \\*
	   Kristina Lerman\\*
       Aram Galstyan\\*
	   Eric Rice & (Outside Member)\\*
       Dana Goldman & (Outside Member)}
       
\begin{preface}
\prefacesection{Abstract}
The field of influence maximization involves finding ``\textit{influential}" members of a social network. This field has seen a lot of progress in recent years, with a significant number of models and algorithms having been proposed in previous work. Unfortunately, most of these models and algorithms do not address challenges that are crucial in many real-world domains such as uncertainties in network structure, selection of nodes in multiple stages, etc. This begets the following question: Can current state-of-the-art influence maximization techniques be deployed successfully in these domains? If not, can influence maximization algorithms, which take into account the aforementioned real-world challenges, be designed? Finally, can these influence maximization algorithms be deployed successfully in the real world for society's benefit? 


This thesis attempts to answer all these questions by focusing on domains where influence maximization can potentially be used for social good. In particular, this thesis is strongly motivated by the following domain: raising awareness about Human Immunodeficiency Virus (HIV) among homeless youth. HIV infections are ten times more common among homeless youth than among youth with stable homes, mainly due to their involvement in high-risk activities such as unprotected sex, sharing drug needles, etc. To prevent spread of HIV, many homeless shelters conduct intervention camps for raising awareness about HIV prevention and treatment practices among homeless youth. However, manpower constraints faced by these shelters mean that they can only intervene upon a small set ($\sim$3-4) of youth in every intervention. Moreover, these shelters prefer a series of  small sized intervention camps organized sequentially. 

With this domain in mind (which is just one of many possible domains), this thesis makes the following technical contributions: (i) we provide a definition of the Dynamic Influence Maximization Under Uncertainty (or DIME) problem, which models the problem faced by homeless shelters accurately; (ii) we propose a novel Partially Observable Markov Decision Process (POMDP) model for solving the DIME problem; (iii) we design two scalable POMDP algorithms (PSINET and HEALER) for solving the DIME problem, since conventional POMDP solvers fail to scale up to sizes of interest; and (iv) we test our algorithms effectiveness in the real world by conducting a pilot study with actual homeless youth in Los Angeles. The success of this pilot (as explained later) shows the promise of using influence maximization for social good on a larger scale.

Further, we also conduct extensive experimentation by simulating our algorithms on (i) real-world networks of homeless youth; and (ii) artificially generated networks like Erdos-Renyi, Watts-Strogatz, etc. Our results show that our POMDP algorithms (i) are able to successfully scale up to real-world network sizes; (ii) are robust to changes in input parameter values; and (iii) our algorithm's performance over baselines improves as network sizes scale up. 

In the future, we plan to improve the ease of our system's adoption by homeless shelter officials by designing an explanation system for justifying solutions generated by our algorithms in an intuitive manner. We also plan to explore hybrid approaches (i.e., combination of POMDPs and greedy methods) to solve the DIME problem. Finally, we also plan to conduct a much larger study with 900 homeless youth to further validate the effectiveness of our algorithms.
\end{preface}

\chapter{Introduction}
Since ancient times, humans have intertwined themselves into various social networks. These networks can be of many different kinds, such as friendship based networks, consumer/seller networks, professional networks, etc. Besides these networks being used for more direct reasons (e.g., friendship based networks used for connecting with old and new friends, consumer/seller networks used for finding out highly rated sellers providing cheap deals, etc.), these networks also play a critical role in the formulation and propagation of opinions and ideas among the people in that network. In recent times, this property of social networks has been exploited by companies to popularize their products among consumers, and by election campaigners to win support among voters, among other examples.

From a computer science perspective, the question of finding out the most ``\textit{influential}" people in a social network is well studied in the field of influence maximization, which looks at the problem of selecting the $K$ (an input parameter) most influential nodes in a social network (represented as a graph), who will be able to influence the most number of people in the network within a given time period. Influence in these networks is assumed to spread according to a known \textit{influence model} (popular ones are independent cascade \cite{leskovec2007cost} and linear threshold \cite{chen2010scalable}). Since the field's inception in 2003 by Kempe et. al. \cite{kempe2003maximizing}, influence maximization has seen a lot of progress over the years \cite{leskovec2007cost, kimura2006tractable, chen2010scalable, cohen2014sketch, Borgs14, tang2014influence, bharathi2007competitive, kostka2008word, borodin2010threshold, lerman2016majority,ghosh2009leaders,ghosh2010community,ver2013information,galstyan2009maximizing,galstyan2008influence}.

Unfortunately, most models and algorithms from previous work do not address challenges that are crucial in many real-world domains such as uncertainties in network structure, selection of nodes in multiple stages, etc. Specifically, most previous work suffers from four potential major limitations. First, almost every previous work focuses on single-shot decision problems, where only a single subset of graph nodes is to be chosen and then evaluated for influence spread. Instead, most realistic applications of influence maximization would require selection of nodes in multiple stages (as we illustrate later). Second, the state of the network is not known; thus, the selection of nodes in multiple stages (which is unhandled in previous work) introduces additional uncertainty about which network nodes are influenced at a given point in time, which complicates the node selection procedure. Third, network structure is assumed to be known with certainty in most previous work, which is untrue in reality, considering that there is always noise in any network data collection procedure. Fourth, most previous work in influence maximization has used two probabalistic influence models (the independent cascade model and the linear threshold model) for their mathematical elegance, but their effectiveness in modeling the influence spread process remains invalidated in the real world. 


This thesis attempts to resolve these limitations and is motivated by an important domain, where influence maximization could be used for social good: raising awareness about Human Immunodeficiency Virus (HIV) among homeless youth. HIV-AIDS is a dangerous disease which claims 1.5 million lives annually \cite{unaids}, and homeless youth are particularly vulnerable to HIV due to their involvement in high risk behavior such as unprotected sex, sharing drug needles, etc. \cite{nchc}. To prevent the spread of HIV, many homeless shelters conduct intervention camps, where a select group of homeless youth are trained as ``peer leaders" to lead their peers towards safer practices and behaviors, by giving them information about safe HIV prevention and treatment practices. These peer leaders are then tasked with spreading this information among people in their social circle. 

However, due to financial/manpower constraints, the shelters can only organize a limited number of intervention camps. Moreover, in each camp, the shelters can only manage small groups of youth ($\sim$3-4) at a time (as emotional and behavioral problems of youth makes management of bigger groups difficult). Thus, the shelters prefer a series of small sized camps organized sequentially \cite{rice2012}. As a result, the shelter cannot intervene on the entire target (homeless youth) population. Instead, it tries to maximize the spread of awareness among the target population (via word-of-mouth influence) using the limited resources at its disposal. To achieve this goal, the shelter uses the friendship based social network of the target population to strategically choose the participants of their limited intervention camps. Unfortunately, the shelters' job is further complicated by a lack of complete knowledge about the social network's structure \cite{rice2010positive}. Some friendships in the network are known with certainty whereas there is uncertainty about other friendships. 

Thus, the shelters face an important challenge: they need a sequential plan to choose the participants of their sequentially organized interventions. This plan must address three key points: (i) it must deal with network structure uncertainty;  (ii) it needs to take into account new information uncovered during the interventions, which reduces the uncertainty in our understanding of the network; and (iv) the intervention approach should address the challenge of gathering information about social networks of homeless youth, which usually costs thousands of dollars and many months of time \cite{rice2012}. 

The contributions of this thesis are as follows. First, we use the homeless youth domain to motivate the definition of the Dynamic Influence Maximization Under Uncertainty (DIME) problem \cite{yadav2016using,yadav-explain,yadav-ideas,yadav2015handling,yadav2015preventing,yadav2016psinet,yadav2017explanation,yadav2017ibm,yadav2017influence,yadav2017maximizing,yadav2018ijcai}, which models the aforementioned challenge faced by the homeless shelters accurately. Infact, the sequential selection of network nodes in multiple stages in DIME sets it apart from any other previous work in influence maximization \cite{leskovec2007cost, kimura2006tractable, chen2010scalable, cohen2014sketch}. Second, we introduce a novel Partially Observable Markov Decision Process (POMDP) based model for solving DIME, which takes into account uncertainties in network structure and evolving network state . Third, since conventional POMDP solvers fail to scale up to sizes of interest (our POMDP had $2^{300}$ states and ${150 \choose 6}$ actions), we design two scalable (and more importantly, ``\textit{deployable}") algorithms, which use our POMDP model to solve the DIME problem \cite{yadav2018}. 

Our first algorithm PSINET \cite{yadav2015preventing} relies on the following key ideas: (i) compact representation of transition probabilities to manage the intractable state and action spaces; (ii) combination of the QMDP heuristic with Monte-Carlo simulations to avoid exhaustive search of the entire belief space; and (iii) voting on multiple POMDP solutions, each of which efficiently searches a portion of the solution space to improve accuracy. Unfortunately, even though PSINET was able to scale up to real-world sized networks, it completely failed at scaling up in the number of nodes that get picked in every round (intervention). To address this challenge, we designed HEAL, our second algorithm. HEAL \cite{yadav2016using} hierarchically subdivides our \textit{original POMDP} at two layers: (i) In the top layer, graph partitioning techniques are used to divide the \textit{original POMDP} into \textit{intermediate POMDPs}; (ii) In the second level, each of these \textit{intermediate POMDPs} is further simplified by sampling uncertainties in network structure repeatedly to get \textit{sampled POMDPs}; (iii) Finally, we use aggregation techniques to combine the solutions to these simpler POMDPs, in order to generate the overall solution for the \textit{original POMDP}.
 
Finally, we have also tested HEAL's performance in a real-world pilot study, in collaboration with Safe Place for Youth\footnote{\label{ftnote1}http://safeplaceforyouth.nationbuilder.com/}, a homeless shelter in Los Angeles which provides food and lodging to homeless youth. We enrolled 60 homeless youth from Safe Place for Youth and conducted multiple interventions and post-intervention surveys to evaluate the effectiveness of HEAL at spreading awareness about HIV. Our pilot results showed that HEAL was able to propagate information to 66\% of all surveyed youth. More importantly, HEAL was successful in behavior change as well, with a 25\% increase in the number of youth who get tested for HIV regularly. These results point to the usability of HEAL in the real world. More importantly, it illustrates one way (among many others) in which social networks and influence maximization can be harnessed for doing social good.

\chapter{Background and Related Work}
\section{Related Work}
The influence maximization problem, as stated by Kempe et. al. \cite{kempe2003maximizing}, takes in a social network as input (in the form of a graph), and outputs a set of $K$ `\textit{seed nodes}' which maximize the expected influence spread in the social network within $T$ time steps. Note that the expectation of influence spread is taken with respect to a probabilistic influence model (explained later), which is also provided as input to the problem.  

There are many algorithms for finding `\textit{seed sets}' of nodes to maximize influence spread in networks \cite{kempe2003maximizing,leskovec2007cost,Borgs14,tang2014influence,lerman2016majority,ghosh2009leaders,ghosh2010community,ver2013information,galstyan2009maximizing,galstyan2008influence}. However, all these algorithms assume \textit{no uncertainty in the network structure} and select a single seed set. In contrast, we select several seed sets sequentially in our work to select intervention participants, as that is a natural requirement arising from our homeless youth domain. Also, our work takes into account uncertainty about the network structure and influence status of network nodes (i.e., whether a node is influenced or not). Finally, unlike most previous work \cite{kempe2003maximizing,leskovec2007cost,Borgs14,tang2014influence,lerman2016majority,ghosh2009leaders,ghosh2010community,ver2013information,galstyan2009maximizing,galstyan2008influence}, we use a different influence model as we explain later.

There is another line of work by Golovin et. al. \cite{golovin2011adaptive}, which introduces adaptive submodularity and discusses adaptive sequential selection (similar to our problem). They prove that a Greedy algorithm provides a $(1-1/e)$ approximation guarantee. However, unlike our work, they assume no uncertainty in network structure. Also, while our problem can be cast into the adaptive stochastic optimization framework of \cite{golovin2011adaptive}, our influence function is not adaptive submodular (as shown later), because of which their Greedy algorithm loses its approximation guarantees. 

\section{Network Representation and Influence Model}
We represent social networks as directed graphs (consisting of \textit{nodes} and \textit{directed edges}) where each \textit{node} represents a person in the social network and a \textit{directed edge} between two nodes $A$ and $B$ (say) represents that node $A$ \textit{considers} node $B$ as his/her friend. \textit{We assume directed-ness of edges as sometimes homeless shelters assess that the influence in a friendship is very much uni-directional; and to account for uni-directional follower links on Facebook}. Otherwise friendships are encoded as two uni-directional links.
 
\textbf{Uncertain Network}: The uncertain network is a directed graph $G=(V,E)$  with $|V| = N$ nodes and $|E| = M$ edges. The edge set $E$ consists of two disjoint subsets of edges: $E_c$ (the set of certain edges, i.e., friendships which we are certain about) and $E_u$ (the set of uncertain edges, i.e., friendships which we are uncertain about). Note that uncertainties about friendships exist because HEALER's Facebook application misses out on some links between people who are friends in real life, but not on Facebook.

\begin{figure}[t]
\center{\includegraphics[scale=.35]
{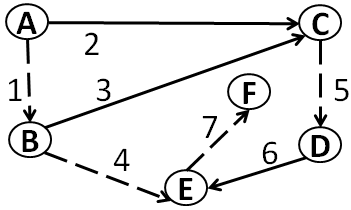}}
\caption{\label{fig:uncertainG} Uncertain Network}
\vspace{-2mm}
\end{figure}

To model the uncertainty about missing edges, every uncertain edge $e \in E_u$ has an existence probability $u(e)$ associated with it, which represents the likelihood of ``existence" of that uncertain edge. For example, if there is an uncertain edge $(A,B)$ (i.e., we are unsure whether node $B$ is node $A$'s friend), then $u(A,B) = 0.75$ implies that $B$ is $A$'s friend with a 0.75 chance. In addition, each edge $e \in E$ (both certain and uncertain) has a propagation probability $p(e)$ associated with it. A propagation probability of 0.5 on directed edge $(A,B)$ denotes that if node $A$ is influenced (i.e., has information about HIV prevention), it influences node $B$ (i.e., gives information to node $B$) with a 0.5 probability in each subsequent time step (our full influence model is defined below). This graph $G$ with all relevant $p(e)$ and $u(e)$ values represents an uncertain network and serves as an input to the DIME problem. Figure \ref{fig:uncertainG} shows an uncertain network on 6 nodes (\textit{A} to \textit{F}) and 7 edges. The dashed and solid edges represent uncertain (edge numbers 1, 4, 5 and 7) and certain (edge numbers 2, 3 and 6) edges, respectively.

\textbf{Influence Model} We use a variant of the independent cascade model \cite{yan2011influence}. In the standard independent cascade model, all nodes that get influenced at round $t$ get a \textbf{single} chance to influence their un-influenced neighbors at time $t+1$. If they fail to spread influence in this \textbf{single} chance, they don't spread influence to their neighbors in future rounds. Our model is different in that we assume that nodes get \textbf{multiple} chances to influence their un-influenced neighbors. If they succeed in influencing a neighbor at a given time step $t'$, they stop influencing that neighbor for all future time steps. Otherwise, if they fail in step $t'$, they try to influence again in the next round. This variant of independent cascade has been shown to empirically provide a better approximation to real influence spread than the standard independent cascade model \cite{cointet2007, yan2011influence}. Further, we assume that nodes that get influenced at a certain time step remain influenced for all future time steps. 

\section{POMDP}
Partially Observable Markov Decision Processes (POMDPs) are a well studied model for sequential decision making under uncertainty \cite{puterman2009markov}. Intuitively, POMDPs model situations wherein an agent tries to maximize its expected long term \textit{rewards} by taking various \textit{actions}, while operating in an environment (which could exist in one of several \textit{states} at any given point in time) which reveals itself in the form of various \textit{observations}. The key point is that the exact state of the world is not known to the agent and thus, these actions have to be chosen by reasoning about the agent's probabilistic beliefs (belief state). The agent, thus, takes an action (based on its current belief), and the environment transitions to a new world state. However, information about this new world state is only partially revealed to the agent through observations that it gets upon reaching the new world state. Hence, based on the agent's current belief state, the action that it took in that belief state, and the observation that it received, the agent updates its belief state. The entire process repeats several times until the environment reaches a terminal state (according to the agent's belief).

More formally, a full description of the POMDP includes the sets of possible environment \textit{states}, the set of \textit{actions} that the agent can take, and the set of possible \textit{observations} that the agent can observe. In addition, the full POMDP description includes a \textit{transition matrix}, for storing \textit{transition probabilities}, which specify the probability with which the environment transitions from one \textit{state} to another, conditioned on the immediate \textit{action} taken. Another component of the POMDP description is the \textit{observation matrix}, for storing \textit{observation probabilities}, which specify the probability of getting different \textit{observations} in different \textit{states}, conditioned on the \textit{action} taken to reach that \textit{state}. Finally, the POMDP description includes a \textit{reward} matrix, which specifies the agent's \textit{reward} of taking \textit{actions} in different \textit{states}.	

A POMDP policy $\Pi$ provides a mapping from every possible belief state (which is a probability distribution over world states) to an action $a=\Pi(\beta)$. Our aim is to find an optimal policy $\Pi^*$ which, given an initial belief $\beta_0$, maximizes the expected cumulative long term reward over H horizons (where the agent takes an action and gets a reward in each time step until the horizon H is reached). Computing optimal policies offline for finite horizon POMDPs is PSPACE-Complete. Thus, focus has recently turned towards online algorithms, which only find the best action for the current belief state \cite{paquet2005online,silver2010monte}. Thus, online planning interleaves planning and execution at every time step.

\chapter{DIME Problem}
We now provide some background information that helps us define a precise problem statement for DIME. After that, we will show some hardness results about this problem statement. 

Given the \textit{uncertain network} as input, we plan to run for \textit{$T$ rounds} (corresponding to the number of interventions organized by the homeless shelter).
In each round, we will choose \textit{$K$ nodes} (youth) as intervention participants. These participants are assumed to be influenced post-intervention with certainty. Upon influencing the chosen nodes, we will `\textit{observe}' the true state of the \textit{uncertain edges} (friendships) out-going from the selected nodes. This translates to asking intervention participants about their 1-hop social circles, which is within the homeless shelter's capabilities \cite{rice2012position}. 

After each round, influence spreads in the network according to our influence model for \textit{$L$ time steps}, before we begin the next round. This $L$ represents the time duration in between two successive intervention camps. \textit{In between rounds, we do not observe the nodes that get influenced during $L$ time steps}. We only know that explicitly chosen nodes (our intervention participants in all past rounds) are influenced. Informally then, given an uncertain network $G_0=(V, E)$ and integers $T$, $K$, and $L$ (as defined above), our goal is to find an online policy for choosing \textit{exactly} $K$ nodes for $T$ successive rounds (interventions) which maximizes influence spread in the network at the end of $T$ rounds.

We now provide notation for defining an online policy formally. Let $\bm{\mathcal{A}}=\{A \subset V \mbox{ s.t. } |A|=K \}$ denote the set of $K$ sized subsets of $V$, which represents the set of possible choices that we can make at every time step $t \in [1,T]$. Let $A_i \in \bm{\mathcal{A}} \mbox{ } \forall i \in [1,T]$ denote our choice in the $i^{th}$ time step. Upon making choice $A_i$, we `\textit{observe}' uncertain edges adjacent to nodes in $A_i$, which updates its understanding of the network. Let $G_i \mbox{ } \forall \mbox{ } i \in [1,T]$ denote the uncertain network resulting from $G_{i-1}$ with \textit{observed} (additional edge) information from $A_i$. Formally, we define a history $H_i \mbox{ } \forall \mbox{ } i \in [1,T]$ of length $i$ as a tuple of past choices and observations $H_i = \tuple{G_0, A_1, G_1, A_2,..,A_{i-1},G_i}$. Denote by $\bm{\mathcal{H}_i} = \{ H_k \mbox{ s.t. } k \leqslant i \}$ the set of all possible histories of length less than or equal to $i$. Finally, we define an $i$-step policy $\bm{\Pi_i} \colon \bm{\mathcal{H}_i} \to \bm{\mathcal{A}}$ as a function that takes in histories of length less than or equal to $i$ and outputs a $K$ node choice for the current time step. We now provide an explicit problem statement for DIME.

\begin{problem}{\textbf{DIME Problem}}
Given as input an uncertain network $G_0=(V, E)$ and integers $T$, $K$, and $L$ (as defined above). Denote by $\mathcal{R}(H_T, A_T)$ the \textit{expected total number of influenced nodes at the end of round $T$}, given the $T$-length history of previous observations and actions $H_T$, along with $A_T$, the action chosen at time $T$. Let $\E_{H_T,A_T \sim \Pi_T} [\mathcal{R}(H_T,A_T)]$ denote the expectation over the random variables $H_T=\tuple{G_0, A_1,..,A_{T-1},G_T}$ and $A_T$, where $A_i$ are chosen according to $\Pi_T(H_i)  \mbox{ } \forall \mbox{ } i \in [1,T]$, and $G_i$ are drawn according to the distribution over uncertain edges of $G_{i-1}$ that are revealed by $A_i$. The objective of DIME is to find an optimal $T$-step policy $\bm{\Pi_T^*} = \argmax_{\Pi_T} \E_{H_T,A_T \sim \Pi_T}[\mathcal{R}(H_T, A_T)]$. 
\end{problem} 

We now analyze the hardness of computation in the DIME problem in the next two theorems. More details and proofs of these theorems can be found in Yadav et. al. \cite{yadav2016using}.

\begin{theorem}\label{Th:1}
The DIME problem is NP-Hard.
\end{theorem}

Some NP-Hard problems exhibit nice properties that enable approximation guarantees for them. Golovin et. al. \cite{golovin2011adaptive} introduced adaptive submodularity, an analog of submodularity for adaptive settings. Presence of adaptive submodularity ensures that a simply greedy algorithm provides a $(1-1/e)$ approximation guarantee w.r.t. the optimal solution defined on the \textit{uncertain network}. However, as we show next, while DIME can be cast into the adaptive stochastic optimization framework of \cite{golovin2011adaptive}, our influence function is not adaptive submodular, because of which their Greedy algorithm does not have a $(1-1/e)$ approximation guarantee. 

\begin{theorem}\label{Th:2}
The influence function of DIME is not adaptive submodular.
\end{theorem}

\chapter{POMDP Model}
The theorems in the previous chapter show that DIME is a hard problem as it is difficult to even obtain any reasonable approximations. We model DIME as a POMDP \cite{puterman2009markov} because of two reasons. First, POMDPs are a good fit for DIME as (i) we conduct several interventions sequentially, similar to sequential POMDP actions; and (ii) we have \textit{partial observability} (similar to POMDPs) due to uncertainties in network structure and influence status of nodes. Second, POMDP solvers have recently shown great promise in generating near-optimal policies efficiently \cite{silver2010monte}. We now 
explain how we map DIME onto a POMDP. 

\textbf{States. } A POMDP state in our problem is a pair of binary tuples $s = \tuple{W, F}$ where $W$ and $F$ are of lengths $|V|$ and $|E_U|$, respectively. Intuitively, $W$ denotes the influence status of network nodes, where $W_i = 1$ denotes that node $i$ is influenced and $W_i = 0$ otherwise. Moreover, $F$ denotes the existence of uncertain edges, where $F_i = 1$ denotes that the $i^{th}$ uncertain edge exists in reality, and $F_i = 0$ otherwise.

\textbf{Actions. } Every choice of a subset of $K$ nodes is a POMDP action. More formally, $A = \{ a \subset V s.t. |a| = K\}$. For example, in Figure \ref{fig:uncertainG}, one possible action is $\{A,B\}$ (when $K=2$).

\textbf{Observations. }Upon taking a POMDP action, we ``\textit{observe}" the ground reality of the uncertain edges outgoing from the nodes chosen in that action. Consider $\Theta(a) = \{ \mbox{e }|\mbox{ e = (x,y) \text{s.t.} x} \in a \mbox{ } \wedge\mbox{ e} \in E_u \}\mbox{ }\forall a \in A$, which represents the (ordered) set of uncertain edges that are observed when we take action $a$. Then, our POMDP observation upon taking action $a$ is defined as $o(a) = \{F_{e} | e \in \Theta(a)\}$, i.e., the F-values of the observed uncertain edges. For example, by taking action $\{B,C\}$ in Figure \ref{fig:uncertainG}, the values of $F_4$ and $F_5$ (i.e., the F-values of uncertain edges in the 1-hop social circle of nodes $B$ and $C$) would be observed.

\textbf{Rewards. } The reward $R(s,a,s')$ of taking action $a$ in state $s$ and reaching state $s'$ is the number of newly influenced nodes in $s'$. More formally, $R(s,a,s') = (\|s'\|-\|s\|)$, where $\|s'\|$ is the number of influenced nodes in $s'$. 

\textbf{Initial Belief State. } The initial belief state is a distribution $\beta_0$ over all states $s \in S$. The support of $\beta_0$ consists of all states $s = \tuple{W, F}$ s.t. $W_i=0 \mbox{ } \forall \mbox{ } i \in [1,|V|]$, i.e., all states in which all network nodes are un-influenced (as we assume that all nodes are un-influenced to begin with). Inside its support, each $F_i$ is distributed independently according to $P(F_i=1)= u(e)$.

\textbf{Transition And Observation Probabilities. } Computation of exact transition probabilities $T(s'|s,a)$ requires considering all possible paths in a graph through which influence could spread, which is $\mathcal{O}(N!)$ ($N$ is number of nodes in the network) in the worst case. Moreover, for large social networks, the size of the transition and observation probability matrix is prohibitively large (due to exponential sizes of state and action space). Therefore, instead of storing huge transition/observation matrices in memory, we follow the paradigm of large-scale online POMDP solvers \cite{silver2010monte,eck2015ask} by using a generative model $\Lambda(s, a) \sim (s', o, r)$ of the transition and observation probabilities. This generative model allows us to generate on-the-fly samples from the exact distributions $T(s'|s,a)$ and $\Omega(o|a,s')$ at very low computational costs. Given an initial state $s$ and an action $a$ to be taken, our generative model $\Lambda$ simulates the random process of influence spread to generate a random new state $s'$, an observation $o$ and the obtained reward $r$. Details of the random process of simulation can be found in Yadav et. al. \cite{yadav2015preventing, yadav2016using}.

First, we tried to solve this POMDP using current state-of-the-art solvers. Initial experiments with ZMDP \cite{zmdp} showed that state-of-the-art offline POMDP planners ran out of memory on 10 node graphs. Thus, we focused on POMCP \cite{silver2010monte}, a state-of-the-art online POMDP solver. Unfortunately, even POMCP ran out of memory on 30 node graphs. In the next two chapters, we provide two different scalable POMDP algorithms (PSINET and HEAL) for solving the DIME problem.

\chapter{PSINET}
As a first attempt at solving DIME, we proposed PSINET, a Monte Carlo (MC) sampling based online planner that simplifies the POMDP into smaller sub-problems and then uses various kinds of voting techniques to aggregate the solutions of the sub-problems.

\begin{figure}[t]
\subfigure[Overall flow of PSINET]{\includegraphics[height=1.8in,width=0.49\columnwidth]{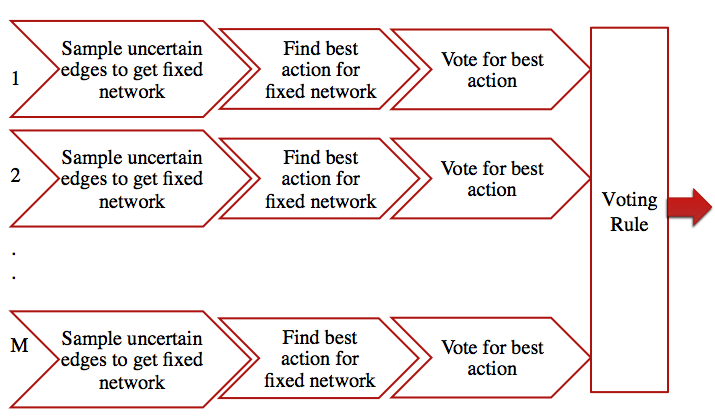}\label{fig:Figure8}}
\hspace{2mm}
\subfigure[Experimental results show improvement over previous work]{\includegraphics[height=1.8in,width=0.40\columnwidth]{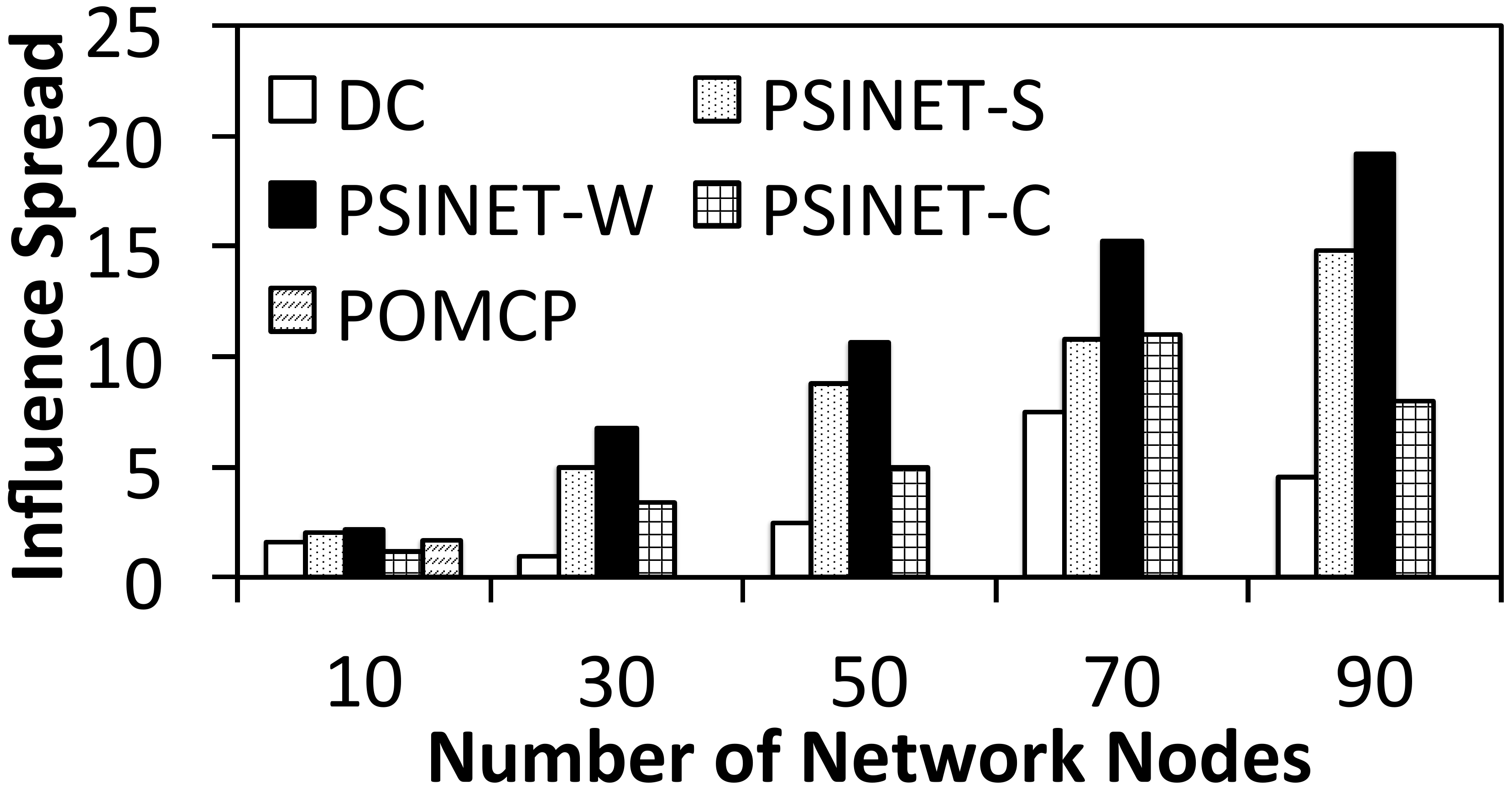}\label{fig:Figure11}}
\caption{\small PSINET Flow and Results}
\end{figure}


The flow of PSINET proceeds in three distinct steps, as shown in Figure \ref{fig:Figure8}. First, graph sampling techniques are used to get many different graph instances, each of which is solved as an independent POMDP. Second, each of these independent POMDPs is run (in parallel) to find out the best action to take in each independent POMDP. Finally, all these independent POMDPs vote for the best action to be taken. These votes are then aggregated using different voting mechanisms (each different voting mechanism gives rise to different variants of PSINET, as we explain later), and the overall best action that wins this vote is executed by PSINET. We now go through each of the three distinct steps shown in Figure \ref{fig:Figure8} in more detail.

\textbf{Sampling Graphs} First, we randomly keep or remove uncertain edges to create one graph instance. As a single instance might not represent the real network well, we instantiate the graph several times to create a set $\mathbf{\Delta}$ of instances. Then, we use each of these instances to vote for the best action to be taken. 
 
\textbf{Finding Best Action} For each sampled graph instance, we find the action which maximizes long term rewards averaged across $\eta$ (we use $\eta$=2\textsuperscript{8}) MC simulations starting from states (particles) sampled from the current belief $\beta$. Each MC simulation samples a particle from $\beta$ and chooses an action to take (choice of action is explained in Yadav et. al. \cite{yadav2015preventing}). Then, upon taking this action, we follow a uniform random rollout policy (until either termination, i.e., all nodes get influenced, or the horizon is breached) to estimate the long term reward, which we get by taking the ``selected" action. Finally, we pick the action with the maximum average reward. 

\textbf{Voting Mechanisms} Finally, each network instance votes for the best action (found in the previous step) to take in the uncertain graph and the overall best action (approximate) is chosen by aggregating these votes according to different voting schemes. We propose using the following three different voting schemes, each of which leads to a different variant of PSINET:

\textbf{PSINET-S} Each instance's vote gets equal weight.

\textbf{PSINET-W} Every instance's vote gets weighted differently. This weighting scheme approximates the probabilities of occurrences of real world events by giving low weights to instances which removes either too few or too many uncertain edges, since those events are less likely to occur. Instances which remove exactly half the number of uncertain edges get the highest weight, since that event is most likely \cite{yadav2015preventing}.

\textbf{PSINET-C} Given a ranking over actions, the Copeland rule makes pairwise comparisons among all actions, and picks the one preferred by a majority of instances over the highest number of other actions \cite{pomerol2000multicriterion}. In PSINET-C, similar to Jiang. et. al. \cite{Jiang2014}, rankings over actions are generated for each instance, and the winner decided using Copeland rule.

Figure \ref{fig:Figure11} compares solution qualities of Degree Centrality (DC), POMCP and PSINET-(S,W and C) on BTER \cite{seshadhri2012community} networks of varying sizes. DC selects nodes in decreasing order of out-degrees, where every uncertain edge \textit{e} adds \textit{u(e)} to the node degrees. We choose DC as our baseline as it is the current modus operandi of agencies working with homeless youth. The x-axis shows number of network nodes and the y-axis shows influence spread across varying horizons (number of interventions). This figure shows that all POMDP based algorithms beat DC by around 60\%, which shows the value of our POMDP model. Further, it shows that PSINET-W beats PSINET-(S and C). Also, \textit{POMCP runs out of memory on 30 node graphs}.


Unfortunately, even though PSINET was able to scale up to real-world sized networks, it completely failed at scaling up in the number of nodes that get picked in every round (intervention). Thus, while PSINET was successful in scaling up to the required POMDP state space, it failed to deal with the explosion in action space that occurred with scale up in the number of nodes picked per round. To address this challenge, we designed HEAL, which we present next.

\chapter{HEAL}
\begin{figure}[t]
\center{\includegraphics[scale=.35]
{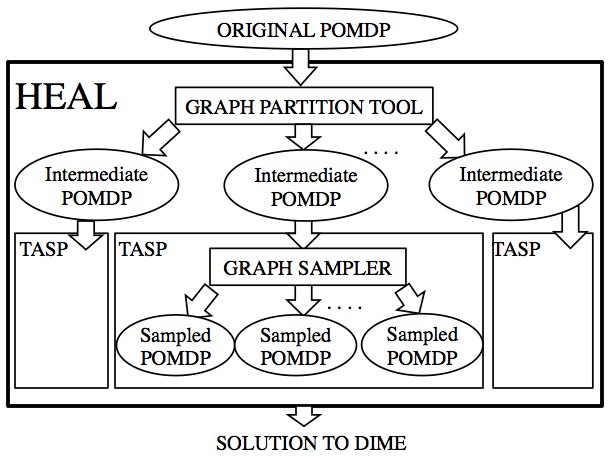}}
\caption{\label{fig:Flow} Hierarchical decomposition in HEAL}
\vspace{-2mm}
\end{figure}

HEAL solves the \textit{original POMDP} using a novel \textit{hierarchical ensembling heuristic}: it creates ensembles of imperfect (and smaller) POMDPs at \textit{two} different layers, in a hierarchical manner (see Figure \ref{fig:Flow}). HEAL's \textit{top layer} creates an ensemble of smaller sized \textit{intermediate POMDPs} by subdividing the original \textit{uncertain network} into several smaller sized \textit{partitioned networks} by using graph partitioning techniques \cite{lasalle2013multi}, which generates partitions that minimize the edges going across partitions (while ensuring that partitions have similar sizes). Since these partitions are ``almost" disconnected, we solve each partition separately. Each of these partitioned networks is then mapped onto a POMDP, and these \textit{intermediate POMDPs} form our \textit{top layer} ensemble of POMDP solvers.

In the bottom layer, each \textit{intermediate POMDP} is solved using TASP (\textbf{T}ree \textbf{A}ggregation for \textbf{S}equential \textbf{P}lanning), our novel POMDP planner, which subdivides the POMDP into another ensemble of smaller sized \textit{sampled POMDPs}. Each member of this \textit{bottom layer} ensemble is created by randomly sampling uncertain edges of the partitioned network to get a sampled network having no uncertain edges, and this sampled network is then mapped onto a \textit{sampled POMDP}. Each sampled POMDP is solved using a novel tree search algorithm, which avoids the exponential branching factor seen in PSINET \cite{yadav2015preventing}. Finally, the solutions of POMDPs in both the \textit{bottom} and \textit{top layer} ensembles are aggregated using novel techniques to get the solution for HEAL's original POMDP. 

These heuristics enable scale up to real-world sizes (at the expense of sacrificing performance guarantees), as instead of solving one huge problem, we now solve several smaller problems. However, these heuristics perform very well in practice. Our simulations show that even on smaller settings, HEAL achieves a 100X speed up over PSINET, while providing a 70\% improvement in solution quality; and on larger problems, \textit{where PSINET is unable to run at all}, HEAL continues to provide high solution quality.

\textbf{Bottom layer: TASP} We now explain TASP, our new POMDP solver that solves each \textit{intermediate POMDP} in HEAL's bottom layer. Given an \textit{intermediate POMDP} and the uncertain network it is defined on, as input, TASP goes through four steps. First, it creates an ensemble of smaller sized \textit{sampled POMDPs}, by sampling uncertain edges of the input network to get an \textit{instantiated} network. Each uncertain edge in the input network is randomly kept with probability $u(e)$, or removed with probability $1-u(e)$, to get an \textit{instantiated} network with no uncertain edges. We repeat this sampling process to get $\Delta$ (a variable parameter) different \textit{instantiated} networks. These $\Delta$ different \textit{instantiated} networks are then mapped onto to $\Delta$ different POMDPs, which form our ensemble of \textit{sampled POMDPs}. Each \textit{sampled POMDP} shares the same action space (defined on the input partitioned network) as the different POMDPs only differ in the sampling of uncertain edges. 

Next, for each instantiated network $\delta \in [1,\Delta]$, we generate an $\alpha^{\delta}$ list of rewards. The $i^{th}$ element of $\alpha^{\delta}$ gives the long term reward achieved by taking the $i^{th}$ action in \textit{instantiated} network $\delta$. After these $\alpha^{\delta}$ lists are generated, we find the expected reward $r_i$ of taking the $i^{th}$ action, by taking a reward expectation across the $\alpha^{\delta}$ lists (for each $\delta \in [1,\Delta]$). For e.g., if $\alpha^{\delta_1}_1 = 10$ and $\alpha^{\delta_2}_1 = 20$, i.e., the rewards of taking the $1^{st}$ action in instantiated networks $\delta_1$ and $\delta_2$ (which occurs with probabilities $P(\delta_1)$ and $P(\delta_2)$) are 10 and 20 respectively, then the expected reward $r_1 = P(\delta_1)\times 10 + P(\delta_2)\times 20$. Note that $P(\delta_1)$ and $P(\delta_2)$ are found by multiplying existence probabilities $u(e)$ (or $1-u(e)$) for uncertain edges that were kept (or removed) in $\delta_1$ and $\delta_2$. Finally, the action $\kappa = \argmax_j r_j$ is returned by TASP. More details on the generation of $\alpha^{\delta}$ lists can be found in Yadav et. al. \cite{yadav2016using}.

\textbf{Top layer: Using Graph Partitioning} We now explain HEAL's top layer, in which we use METIS \cite{lasalle2013multi}, a state-of-the-art graph partitioning technique, to subdivide our original uncertain network into different partitioned networks. These partitioned networks form the ensemble of \textit{intermediate POMDPs} (in Figure \ref{fig:Flow}) in HEAL. Then, TASP (our bottom layer solver) is invoked on each intermediate POMDP independently, and their solutions are aggregated to get the final DIME solution. We try two different partitioning/aggregation techniques, which leads to two variants of HEAL:

\textbf{K Partition Variant (HEAL): } Given the \textit{uncertain} network $G$ and the parameters $K$, $L$ and $T$ as input, we first partition the uncertain network into $K$ partitions. In each round from 1 to $T$, we invoke the bottom layer TASP algorithm to select 1 node from each of the $K$ clusters. These singly selected nodes from the $K$ clusters give us an action of $K$ nodes, which is given to shelter officials to execute. Based on the \textit{observation} (about uncertain edges) that officials get while executing the action, we update the partition networks (which are input to the \textit{intermediate POMDPs}) by either replacing the \textit{observed} uncertain edges with certain edges (if the edge was \textit{observed} to exist in reality) or removing the uncertain edge altogether (if the edge was \textit{observed} to \textit{not exist} in reality). 

\textbf{T Partition Variant (HEAL-T): } Given the \textit{uncertain} network $G$ and the parameters $K$, $L$ and $T$ as input, we first partition the uncertain network into $T$ partitions and TASP picks $K$ nodes from the $i^{th}$ partition ($i \in [1,T]$) in the $i^{th}$ round. 

\begin{figure}[t]
\subfigure[Solution quality]{\includegraphics[height=1.6in,width=0.47\columnwidth]{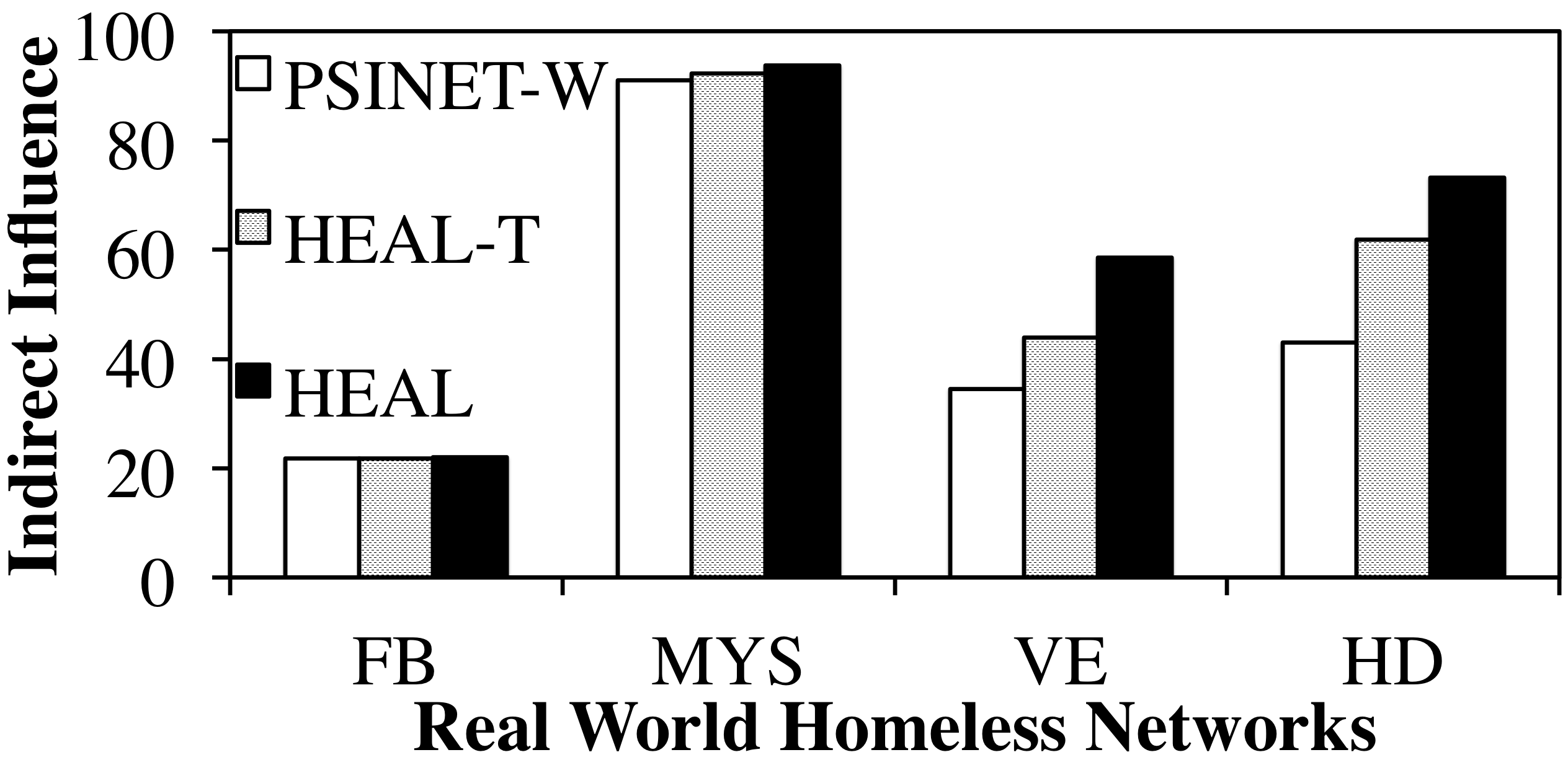}\label{fig:SolQual}}
\hspace{2mm}
\subfigure[Scale up in number of nodes picked per round]{\includegraphics[height=1.6in,width=0.51\columnwidth]{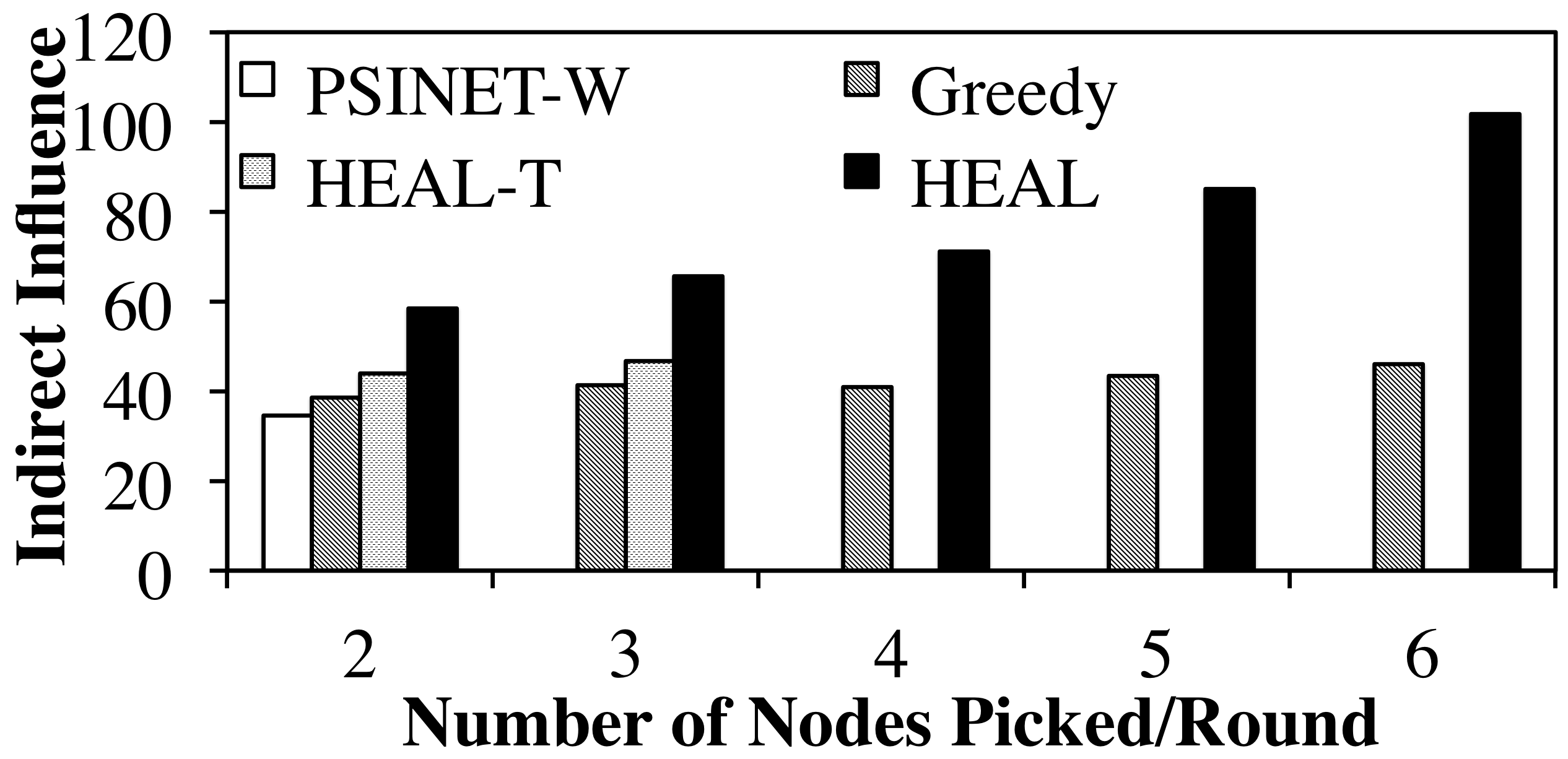}\label{fig:VeniceNet}}
\caption{\small Experimental results show improvement over previous work}
\end{figure}

\textbf{Simulation Results} Figure \ref{fig:SolQual} shows the influence spread of different algorithms on four real world networks of homeless youth. The x-axis shows the four networks and the y-axis shows the influence spread achieved by the different algorithms. This figure shows that (i) HEAL outperforms all other algorithms on every network; (ii) \textit{it achieves $\sim$70\% improvement over PSINET-W} in VE and HD networks; (iii) it achieves $\sim$25\% improvement over HEAL-T. The difference between HEAL and other algorithms is not significant in the FB) and MYS networks, as HEAL is already influencing almost all nodes in these two relatively small networks.

Figure \ref{fig:VeniceNet} shows the influence spread achieved by HEAL, HEAL-T, Greedy and PSINET-W on the VE and HD networks respectively ($T=5$), as we scale up values of $K$, i.e., number of nodes picked per round. The x-axis shows increasing $K$ values, and the y-axis shows the influence spread. This figure show that (i) PSINET-W and HEAL-T fail to scale up  -- they cannot handle more than $K=2$ and $K=3$ respectively (thereby not fulfilling real world demands); (ii) HEAL outperforms all other algorithms, and the difference between HEAL and Greedy increases linearly with increasing $K$ values.

\chapter{Real World Pilot Study}
We conducted a pilot study \cite{rice2018piloting} with actual homeless youth in collaboration with Safe Place for Youth (a homeless shelter in Los Angeles) to test the effectiveness of HEAL in spreading information in a social network. We enrolled 62 homeless youth and constructed a social network using their Facebook contact and follower lists. We conducted three interventions, with four homeless youth selected in every intervention. For each intervention, HEAL would take the social network as input and come up with a recommendation of four network nodes as intervention attendees. 

\begin{figure}[htp]
\center{\includegraphics[scale=.27]
{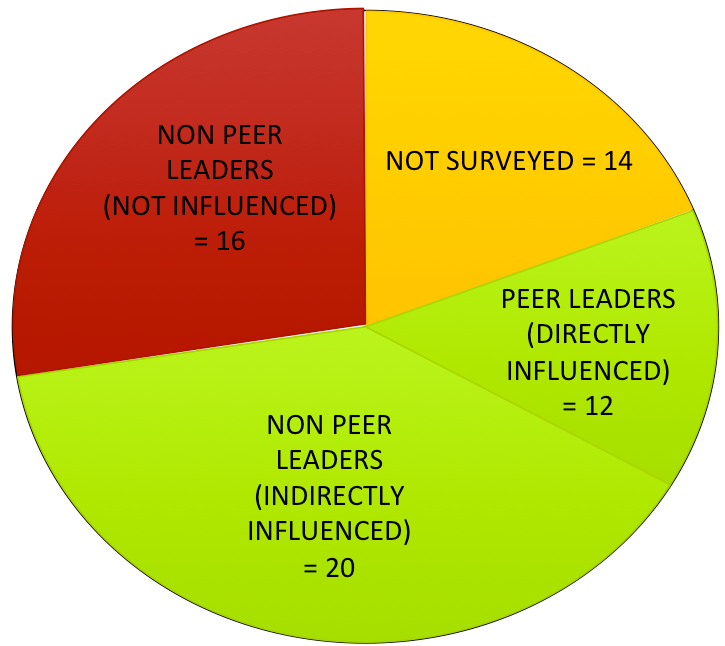}}
\caption{\label{fig:pilot} Result of real-world pilot study}
\vspace{-2mm}
\end{figure}

Figure \ref{fig:pilot} shows that out of 62 homeless youth, 14 youth did not attend the post-intervention surveys. Out of the 48 surveyed youth, 36 youth reported that they received HIV related information from youth in our study. Thus, HEAL was able to influence 66\% of all surveyed youth.

To the best of our knowledge, this is the first deployment of an influence maximization algorithm in the real world. Further, it is the first time that influence maximization has been applied for social good. The success of this pilot study illustrates one way (among many others) in which social networks and influence maximization can be harnessed for doing social good.

\chapter{Future Work}
\section{Explanation system for POMDPs}
The goal of this thesis is to make influence maximization algorithms that actually get deployed in the real world for society's benefit. One big impediment in front of this goal in the homeless youth domain, is the reluctance of homeless shelter officials to adopt state-of-the-art technology (i.e., our algorithms) in selecting their intervention attendees. Thus, there is a need to convince homeless shelter officials that our algorithms are very good at selecting ``\textit{influential}" attendees for their interventions. 

Explanation systems for machine learning algorithms \cite{Tintarev2007ASO} have shown great promise in convincing end-users of the algorithm's usefulness. These explanation systems generate an easy-to-understand explanation of the algorithm's output for the end-user, which facilitates widespread adoption of the algorithm. Recently, explanation systems have also been built for Markov Decision Processes \cite{khan2009minimal,dodson2013english}. However, no previous work has tackled the problem of building explanation systems for POMDPs, due to challenges of handling partial observability in POMDPs. 

In the future, we plan to explore the feasibility of building such an explanation system for HEAL, our POMDP solver. Our goal is to be able to justify the solutions of HEAL to the homeless shelter officials in an intuitive manner. To that end, our first goal is to find out what kind of reasoning do officials (or humans in general) use to pick nodes in very simple graph settings. That will give us an understanding about what kinds of reasons are most likely to persuade humans and offials to adopt HEAL's solutions.

\section{Hybrid approaches}
\begin{figure}[t]
\subfigure[Solution quality]{\includegraphics[height=1.4in,width=0.5\columnwidth]{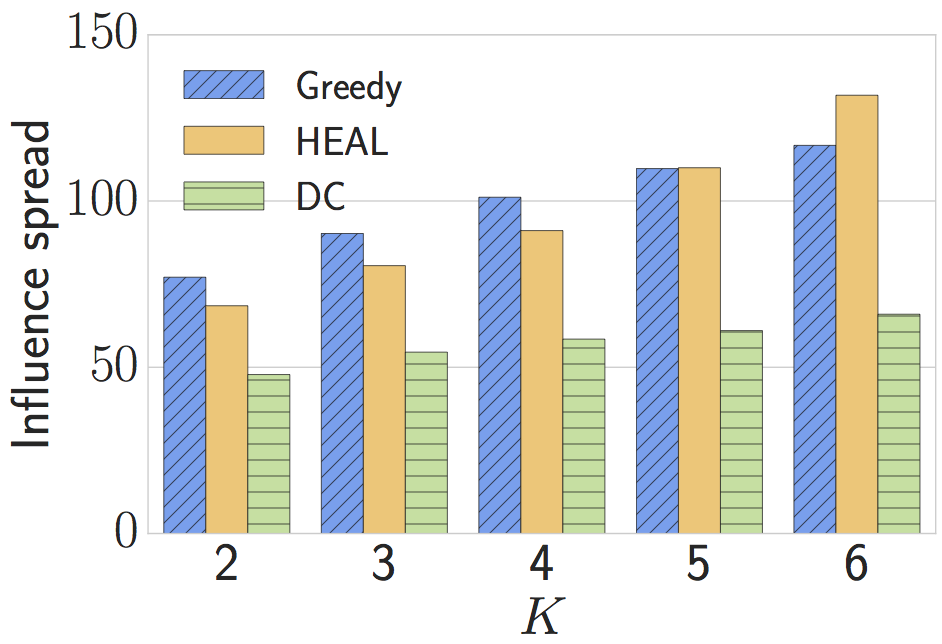}\label{fig:newgreedy}}
\hspace{2mm}
\subfigure[Runtime]{\includegraphics[height=1.4in,width=0.45\columnwidth]{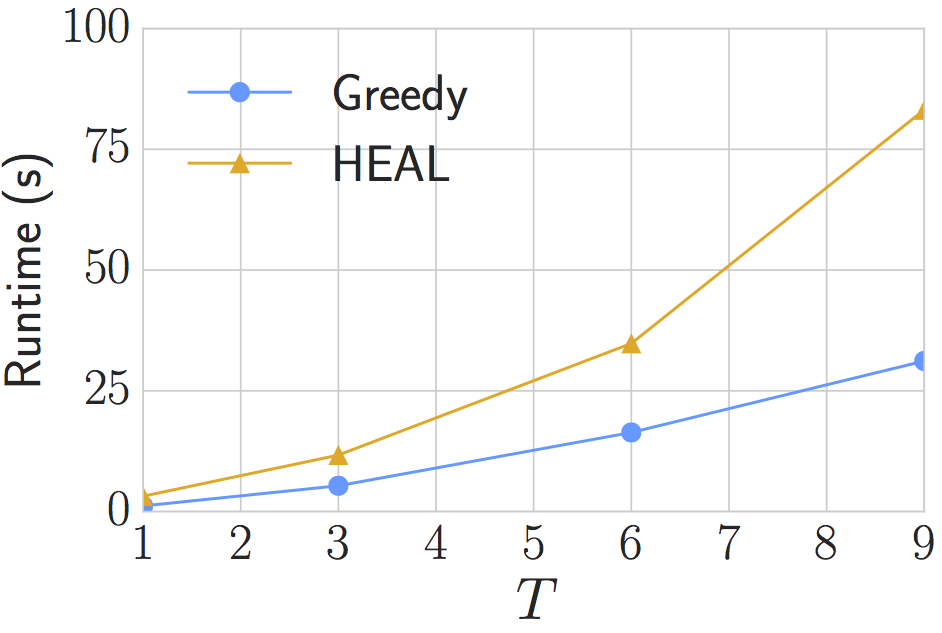}\label{fig:newgreedyruntime}}
\caption{\small New greedy algorithm}
\end{figure}

Recently, we investigated the Greedy algorithm introduced by Golovin et. al. \cite{golovin2011adaptive} more carefully. This work (under submission) was led by another Ph.D. student\footnote{Bryan Wilder was first author in this research}, and resulted in a non-trivial adaptation of Golovin et. al.'s Greedy algorithm \cite{golovin2011adaptive}, which showed promising results. While this new Greedy algorithm has not been tested in the real world, it outperforms our POMDP algorithms (HEAL and PSINET) in simulation for some cases. As shown in Figure \ref{fig:newgreedy}), this Greedy algorithm achieves more influence spread for smaller values of $K$ (i.e., nodes picked per round). It also run much faster than HEAL (as shown in Figure \ref{fig:newgreedyruntime}).

However, there are still cases where HEAL still outperforms this new Greedy algorithm (see large values of $K$ in Figure \ref{fig:newgreedy}). This leads us to the following question: Can we combine ideas from HEAL and the new Greedy algorithm to design a ``\textit{hybrid}" algorithm which runs much faster than HEAL, and outperforms the Greedy algorithm in all cases? We would like to explore this question as part of future work.

\section{Real world validation}
We plan to conduct a much larger study with 900 homeless youth, to test the effectiveness of our algorithms. In this study, we plan to have four different groups of homeless youth (each group has its own social network) which are separated geographically, so as to ensure that the same homeless youth is not a member of two different social networks. We plan to conduct interventions about HIV awareness simultaneously in each of these four groups, with intervention attendees being selected by a different algorithm in each group. This study will allow us to compare the influence spread achieved by different algorithms such as HEAL, Greedy algorithms, Degree Centrality based approaches, uniform random selection, etc. 

Further, the data collected in this study will allow us to validate the effectiveness of different influence spread models (such as independent cascade, linear threshold, or our own repeated activation model, etc.) in the real world. Similar to Lerman et. al. \cite{lerman2012social}, information cascade data about who influenced whom, could be used from this study, which will allow us to fit different influence models to the data, and determine which model provides the best fit.

\begin{singlespace}
\bibliographystyle{plain}
\bibliography{QualifyingExamDocument}
\end{singlespace}

\end{document}